\documentclass[fleqn,usenatbib]{mnras}
\usepackage{hyperref}

\usepackage{newtxtext,newtxmath}
\usepackage[T1]{fontenc}
\usepackage{ae,aecompl}
\usepackage{graphicx}
\usepackage{amsmath}
\usepackage{lipsum}

\usepackage{multirow}
\usepackage{dcolumn}
\usepackage{booktabs}
\usepackage{orcidlink}

\title[Large Volume Tests of a Field Level Emulator]{Learning the Universe: $3\ h^{-1}{\rm Gpc}$ Tests of a Field Level $N$-body Simulation Emulator}

\author[M.~T.~Scoggins et al]{Matthew~T.~Scoggins$^{1}$\thanks{E-mail: mts2188@columbia.edu}\orcidlink{0000-0002-0748-9115},
\ Matthew~Ho$^{1}$,
\ Francisco~Villaescusa-Navarro$^{3,4}$,
\ Drew Jamieson$^{5}$,
\newauthor
\ Ludvig Doeser$^{2}$,
\ and Greg~L.~Bryan$^{1}$
\\
$^{1}$Department of Astronomy, Columbia University, New York, NY, 10027\\
$^{2}$The Oskar Klein Centre, Department of Physics, Stockholm University, AlbaNova University Centre, SE 106 91 Stockholm, Sweden \\
$^{3}$Center for Computational Astrophysics, 162 5th Avenue, New York, NY, 10010, USA \\
$^{4}$Department of Astrophysical Sciences, Princeton University, Peyton Hall, Princeton, NJ 08544, USA\\
$^{5}$Max-Planck-Institut für Astrophysik, Karl-Schwarzschild-Straße 1, 85748 Garching, Germany \\
}

\date{Accepted XXX. Received YYY; in original form ZZZ}

\pubyear{\the\year{}}

\begin{document}
\label{firstpage}
\pagerange{\pageref{firstpage}--\pageref{lastpage}}
\maketitle

\begin{abstract}
    We apply and test a field-level emulator for non-linear cosmic structure formation in a volume matching next-generation surveys. Inferring the cosmological parameters and initial conditions from which the particular galaxy distribution of our Universe was seeded can be achieved by comparing simulated data to observational data. Previous work has focused on building accelerated forward models that efficiently mimic these simulations. One of these accelerated forward models uses machine learning to apply a non-linear correction to the linear $z=0$ Zeldovich approximation (ZA) fields, closely matching the cosmological statistics in the $N$-body simulation. This emulator was trained and tested at $(h^{-1}{\rm Gpc})^3$ volumes, although cosmological inference requires significantly larger volumes. We test this emulator at $(3\ h^{-1}{\rm Gpc})^3$ by comparing emulator outputs to $N$-body simulations for eight unique cosmologies. We consider several summary statistics, applied to both the raw particle fields and the dark matter (DM) haloes. We find that the power spectrum, bispectrum and wavelet statistics of the raw particle fields agree with the $N$-body simulations within ${\sim} 5 \%$ at most scales. For the haloes, we find a similar agreement between the emulator and the $N$-body for power spectrum and bispectrum, though a comparison of the stacked profiles of haloes shows that the emulator has slight errors in the positions of particles in the highly non-linear interior of the halo. At these large $(3\ h^{-1}{\rm Gpc})^3$ volumes, the emulator can create $z=0$ particle fields in a thousandth of the time required for $N$-body simulations and will be a useful tool for large-scale cosmological inference. This is a Learning the Universe publication.
\end{abstract}
\begin{keywords}
cosmological parameters -- machine learning -- large-scale structure of Universe -- haloes
\end{keywords}

\section{Introduction}

The massive stream of observational data arriving from next-generation surveys will revolutionize our ability to distinguish between cosmological theories. However, the defining challenge of modern astronomy will be how we can make best use of these data to maximally extract cosmological information, while overcoming our limited understanding of observational and astrophysical systematics along the way. Classically, perturbative approaches \citep{schmidt2019rigorous, cabass2020eft, ivanov2024full} wrap the effect of these systematics in nuisance parameters, allowing cosmologists to marginalize out this ignorance in exchange for posterior uncertainty. However, these perturbative methods are founded upon linear or quasi-linear growth of large-scale structure, approximations which break down on small scales and in dense regions, where most cosmological information resides \citep{leclercq2021accuracy}. Recently, machine learning models have offered a promising, powerful alternative to perturbative approaches. Instead of defining an explicit likelihood for our observed data, techniques such as implicit inference \citep[also known as simulation-based inference and likelihood-free inference; ][]{cranmer2020frontier, ho2024ltu} learn the relationship between cosmological models and observed data directly from numerical simulations. These simulations can be complex and non-linear, incorporating all cosmological and astrophysical processes which are relevant for the observable at hand.

The drawback of implicit inference approaches is their need for massive suites of simulations to learn from. Machine learning models often need to train on thousands to tens of thousands of examples, varied broadly over the parameters of interest \citep{kaplan2020scaling, ho2024ltu}. Also, each training example must be realistic and representative of a possible observation, which, in a cosmological context, means running a high-fidelity $N$-body or hydrodynamical simulation. Such simulations must not only be high-resolution, to capture observational systematics such as redshift-space distortions and fibre-collisions \citep{modi2025sensitivity, lizancos2025selecting}, but also large volume. Future spectroscopic surveys such as DESI, \textit{Euclid}, and Roman will probe volumes of over $1\ h^{-3}{\rm Gpc}^3$ \citep{eifler2021cosmology, scaramella2022euclid, adame2024desi}, and the simulations will need to realistically couple linear and non-linear structure in large and small scales. Observational systematics such as masks and spectroscopic incompleteness further complicate this behaviour.

Promising alternatives to classical simulations are learned emulators, which accelerate the generation of observable mocks. Emulators trained on a smaller suite of simulations can be expanded to mimic large simulations at more interpolated parameters. Neural network-based emulators can be implemented efficiently on GPUs, massively parallelizing and accelerating the generation of mock training data.
Various emulators have been developed for cosmological summary statistics \citep{euclid2019euclid, bartlett2024syren}, dark matter fields \citep{Jamieson_2023, 2023OJAp....6E..11T, Bartlett_2024}, halo catalogues \citep{Pandey_2024, cuesta2024point, Ding_2024}, baryonic fields \citep{li2021ai, arico2021bacco}, and galaxies \citep{bourdin2024inpainting, nguyen2024dreams}. However, no field-level emulators have been thoroughly tested on the huge volumes for modern cosmological surveys. 

Here, we focus on testing the large volume extrapolation of the $N$-body simulation emulator introduced in \citet{Jamieson_2023}. In this implementation, the Lagrangian displacements and velocities of particles in a dark-matter-only simulation are emulated using two deep convolutional neural networks. The emulator has been integrated into hybrid Lagrangian bias modelling of the galaxy field \citep{Pellejero2024} and Bayesian inference of cosmological initial conditions \citep{Doeser2024}. Extensions of this emulator include training with redshift dependence \citep{Jamieson2024} and with modified gravity \citep{Saadeh2024}. The summary statistics in \citet{Jamieson_2023} show that the emulator's outputs are within ${\sim} 5\%$ of the $N$-body outputs for most scales (typically up to $k{\sim} 0.3$ Mpc$^{-1}h$)  when trained and used on a $(1\ h^{-1}{\rm Gpc})^3$ volume. The main focus of the current paper is to test the performance of this emulator on larger volumes. It is not clear if the emulator can make accurate predictions when scaling to larger volumes, where the emulator will be exposed to modes that are much larger than the network's field-of-view and experience sub-volumes with significantly more variation than the $(1\ h^{-1}{\rm Gpc})^3$ training simulations. To test this, we simulate eight cosmologies at this larger $(3\ h^{-1}{\rm Gpc})^3$ volume, run the emulator, and calculate several summary statistics to measure performance on these larger scales. If successful, the emulator will be an efficient tool for generating observable mocks to be used in implicit inference, which is a primary goal of the Learning the Universe collaboration\footnote{\href{https://learning-the-universe.org/}{learning-the-universe.org}}.

This paper is organized as follows. In \S~\ref{sec:methods} we briefly describe our simulation setup, halo finding procedure, and summarize the emulator. In \S~\ref{sec:results} we present the summary statistics that we use to compare the $N$-body to the emulator, comparing both the particle fields and the haloes. Finally, we summarize our findings and offer our conclusions in~\S~\ref{sec:conclusion}.

\section{Methods}
\label{sec:methods}

\subsection{Simulations and Halo Finding}

In order to test the accuracy and precision of our emulator, we run a set of eight simulations. These simulations follow the evolution of $1536^3$ dark matter particles in a periodic box of $3~h^{-1}{\rm Gpc}$ comoving size. The simulations were run with Gadget-3, an updated version of Gadget-2 \citep{gadget2, gadget3}. The value of the cosmological parameters for these simulations are varied using a Sobol Sequence with priors:
\begin{eqnarray}
0.1\leq\Omega_{\rm m}\leq0.5\\
0.02\leq\Omega_{\rm b}\leq0.08\\
0.5\leq h \leq 0.9\\
0.8\leq n_s \leq 1.2\\
0.6 \leq \sigma_8 \leq 1
\end{eqnarray}
See Table \ref{tab:table1} for the exact value of the parameters. The initial conditions were generated at $z=127$ using second-order Lagrangian Perturbation Theory (2LPT). All work shown is based on the $z=0$ output of these simulations. Dark matter haloes are identified by running the \texttt{ROCKSTAR} halo finder \citep{rockstar}, which applies an algorithm to the 6-D phase space information of the particles (position and velocity) and eliminates unbound particles from the haloes' final measurements.

\begin{table}
    \centering
    \begin{tabular}{ccccccc}
        \toprule
        ID & $\Omega_{\rm m}$ & $\Omega_{\rm b}$ & $ \Omega_{\Lambda}$ & $ h$ & $ \sigma_8$ & $ n_s$ \\
        \midrule
         0  &  0.2823  &  0.0268  &  0.7177  &  0.7165  &  0.9734  &  0.8434 \\
        1  &  0.3699  &  0.0738  &  0.6301  &  0.6873  &  0.7929  &  1.1231 \\
        2  &  0.4054  &  0.0442  &  0.5946  &  0.8973  &  0.6872  &  0.9904 \\
        3  &  0.1429  &  0.0571  &  0.8571  &  0.5049  &  0.8536  &  1.0702 \\
        4  &  0.1965  &  0.0416  &  0.8035  &  0.6072  &  0.8324  &  1.0224 \\
        5  &  0.4592  &  0.0588  &  0.5408  &  0.7902  &  0.6019  &  0.9426 \\
        6  &  0.3157  &  0.0293  &  0.6843  &  0.5755  &  0.7212  &  1.1694 \\
        7  &  0.2283  &  0.0722  &  0.7717  &  0.8204  &  0.9377  &  0.8897 \\
        \bottomrule
    \end{tabular}
    \caption{The values of the cosmological parameters for our 8 simulations. Each set of parameters is generated via a Sobol Sequence.}
    \label{tab:table1}
\end{table}

\subsection{Summary of the emulator}
The emulator we apply in this work is composed of two CNNs that are trained on redshift $z=0$ snapshots from 1,757 of the 2,000 simulations in the Quijote Latin hypercube $N$-body suite \citep{Villaescusa-Navarro_2022}. Each training simulation is run with Gadget-3 \citep{Springel_2005} and has a unique combination of cosmological parameters, ($\Omega_{\rm m}$, $\Omega_b$, $\sigma_8$, $n_s$, $h$). Each simulation consists of $512^3$ particles in a $1~h^{-1}{\rm Gpc}$ cube, and all simulations have different seeds which guarantee unique initial conditions.

If an initial configuration of $N$-body particles is uniformly distributed with positions $\textbf{q}$, and their final positions and velocities at $z=0$, estimated via the ZA or 2LPT, are given by $x(\textbf{q}) = \textbf{q} + \Psi(\textbf{q})$ and $v(\textbf{q}) = \dot{\Psi}(\textbf{q})$, then the CNNs compute a non-linear correction on  the linear estimates of $\Psi(\textbf{q})$ and $\dot{\Psi}(\textbf{q})$. The emulator contains parameters that encode the $\Omega_{\rm m}$ dependence at every layer, and we input this for every cosmology we test. For an in-depth discussion regarding the setup of the emulator, see \cite{Jamieson_2023}.

We test the emulator on 8 different cosmologies, at a volume of $(3\ h^{-1}{\rm Gpc})^3$ with a particle resolution of $1536^3$. In order to run the emulator on these larger volumes, the emulator crops this space into several sub-volumes according to user-defined controls. We choose a crop size of $128^3$ particles, corresponding to a sub-volume of ${\sim} (250\ h^{-1}{\rm Mpc})^3$. This crop size is large enough to capture the large scale structure of the region but still small enough to fit into memory. For more details on this cropping procedure, see the network and training section of \citet{Jamieson_2023}.

While the emulator is capable of running on GPUs, we run the emulator on CPUs which gives a direct timing comparison with $N$-body simulations. The combination of the ZA and the emulator takes less than $10$ CPU hours to complete for a $(3\ h^{-1}{\rm Gpc})^3$ run. This is a massive advantage over the equivalent $N$-body simulation, which took ${\sim} 9$k CPU hours, and even other simulation approximations such as particle mesh simulations which could take up to 150 CPU hours for a similar particle field output \citep{Pandey_2024}. These speed increases represent lower bounds, where a GPU run would result in even more computational efficiency.

\section{Results}
\label{sec:results}
We compare summary statistics between the emulator and the $N$-body outputs, measuring summary statistics on both the particle fields and the halo catalogue generated through \texttt{Rockstar}. We calculate the power spectrum, bispectrum and the wavelet statistics for the particle fields. For the haloes, we compare the mass-weighted power spectrum, mass-weighted bispectrum, halo mass function, and the stacked profiles of haloes in the $N$-body and the emulator for $500$ haloes across different target masses and cosmologies.

In most figures, we compare both the absolute value of the statistics along with the fractional residual between the emulator and the phase-matched $N$-body statistics, where for a given $N$-body statistic $s_{\rm N}$ and emulator statistic $s_{\rm E}$, the fractional residual is always calculated as $\Delta s/s_{\rm } = (s_{\rm E}-s_{\rm N})/s_{\rm N}$.

\subsection{Particles}

\begin{figure}
    \centering
    \includegraphics[width=\columnwidth]{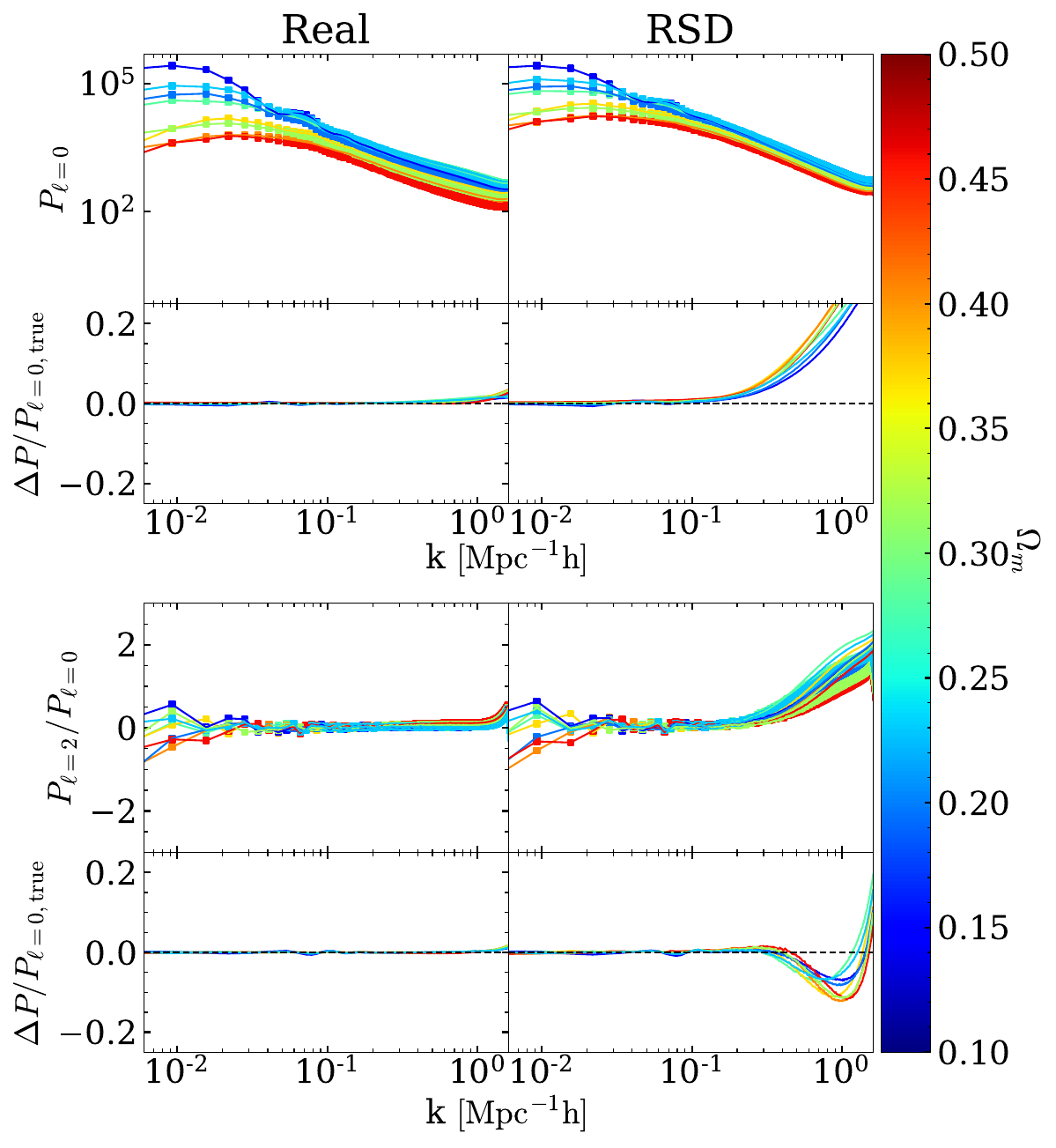}
    \caption{Top: We show the $N$-body (solid) and Emulator (square) outputs of the  monopole ($\ell =0$) matter power spectrum applied to the particle fields. We include both real (left) and RSD (right) outputs. Below each panel, we show the fractional residual error between the $N$-body and the emulator output. Bottom: We show the same as the top, but for the quadrupole ($\ell = 2$) of the power spectrum. Each colour corresponds to one value of $\Omega_{\rm m}$ in our eight cosmologies, with the value shown in the colour bar to the right. Each calculation was performed on a density field created with the CIC mass assignment scheme on a grid of $1536^3$ cells.}
    \label{fig:Pk}
\end{figure}

\textbf{Power Spectrum:} We compute both the real and redshift space distortion (RSD) power spectrum using \texttt{Pylians} \citep{Pylians}. We use \texttt{Pylians}'s cloud-in-cell (CIC) particle mesh assignment scheme to assign the particles to a density field, $n(\boldsymbol{x})$, with $1536^{3}$ cells. We can then compute the over-density $\delta(\boldsymbol{x}) = \frac{n(\boldsymbol{x})}{\overline{n}}  - 1$, where $\overline{n}$ is the mean number of particles per cell. \texttt{Pylians} then Fourier transforms this over-density field, yielding the density modes $\delta(\boldsymbol{k})$, and computes the density power spectrum P(k), following
\begin{equation}
    \langle \delta(\boldsymbol{k})\delta(\boldsymbol{k'})\rangle = (2\pi)^3 \delta^3_{\rm D}(\boldsymbol{k}+\boldsymbol{k'})P(k)
\end{equation}
where $\delta^3_{\rm D}$ is the 3-dimensional Dirac delta function. This can can be expanded into multipole moments,
\begin{equation}
    P(\boldsymbol{k}) = \sum_\ell (2\ell+1) P_{\ell}(k)\mathcal{L}_{\ell}({\rm cos}(\theta))
\end{equation}
for Legendre polynomials $\mathcal{L}_{\ell}$ and angle $\theta$ between the line of sight direction and wave vector $\boldsymbol{k}$. We calculate $P(k)$ for $k_{\rm{min}}=0.003$ Mpc$^{-1}h$ up to $k_{\rm max} = 2.78$ Mpc$^{-1}h$.

In Fig. \ref{fig:Pk} we show the real space 3D and the RSD power spectrum $P_{\ell}(k)$ for poles $\ell= 0,2$. The top row compares the absolute outputs, with the emulator values shown with square markers and the $N$-body outputs shown with solid lines. The bottom row shows the fractional residual, with colours indicating the value of $\Omega_{\rm m}$ for that cosmology. 
Both poles $\ell =0$ and $\ell =2$ show the real power spectrum outputs within ${\sim} 2\%$ of each other across all scales of $k$. We find similar results for RSD outputs, though the emulator and the $N$-body start to diverge as $k \gtrsim 0.3$ Mpc$^{-1}h$. At these smaller scales, the monopole of the RSD power spectrum tends to be overestimated in the emulator, and for the quadrupole, the emulator tends to underestimate these scales. These small scale errors in RSD are due to errors in the emulator's velocity estimates, though bulk large scale velocities are correct. While this is expected from an emulator, a more recent version of this emulator, which is trained on several different redshifts, finds improvements in velocity estimation \citep{Jamieson2024}. Further, the RSD errors are compared to the $N$-body velocities, which are inaccurate for observational RSDs of galaxies. The line-of-sight smearing of the small scale velocities is extremely non-linear and sensitive to baryonic physics. In practice, this must either be modelled with nuisance parameters \citep{Ibanez_2024} or mitigated using selection strategies that avoid the issue \citep{Lizancos_2025}. Either way, the full $N$-body evolution is not a reliable model for RSD observations, so the errors present here are not significant for inference. We also not that the parameter ranges explored here are enormous compared to Planck bounds, where a value of $\Omega_{\rm m}=0.5$ is ${\sim} 26\sigma$ \citep{Planck_2018}. If we restricted these simulations to be within $5 \sigma$ of the most recent Planck results, the errors would improve. \\

\begin{figure*}
    \centering
    \includegraphics[width=1.8\columnwidth]{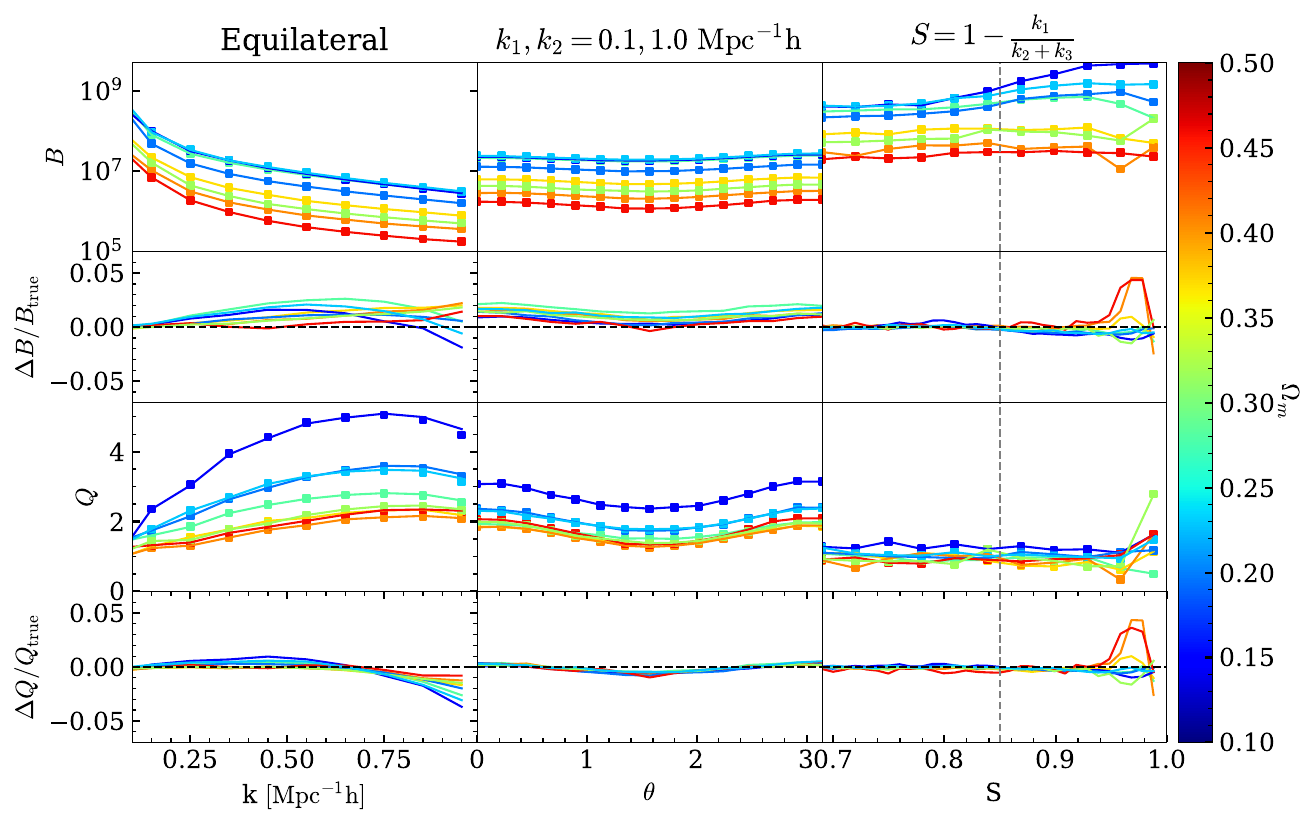}
    \caption{
    We show the $N$-body (solid) and Emulator (square) outputs of the bispectrum $B$ (top two rows) and reduced bispectrum $Q$ (bottom two rows) applied to the particle fields. Each column highlights a unique bispectrum measurement, where the first columns show an equilateral configuration with $k_1=k_2=k_3$ and $k_n \ \epsilon [0.1,0.9]$ Mpc$^{-1}h$. The middle column shows a configuration sweeping over $\theta$ with fixed $k_1$ and $k_2$, with $k_1 = 0.1$ Mpc$^{-1}h$, $k_2 = 1.0$ Mpc$^{-1}h$, and $\theta \ \epsilon [0, \pi]$. Finally, the last column shows the squeezed bispectrum, with parameter $S = 1- \frac{k_1}{k_2+k_3}$ where we have set $k_2 = k_3 =0.1 \ h^{-1}$ Mpc and $k_1$ ranges from $0.0024 \ h^{-1}$ Mpc, slightly larger than the fundamental wave number, up to $ 0.062  \ h^{-1}$ Mpc. The vertical dashed lines show the value of $S$ when $k_1$ is set to the field of view of the emulator, L$_{\rm FOV} \sim 200   \ h^{-1}$ Mpc, or $k_{\rm FOV} \sim 0.03  \ h^{-1}$ Mpc.  Each colour corresponds to one value of $\Omega_{\rm m}$ in our eight cosmologies, with the value shown in the right colour bar. Each calculation was performed on a density field created with the CIC mass assignment scheme on a grid of $1536^3$ cells.
    }
    \label{fig:Bk}
\end{figure*}

\noindent \textbf{Bispectrum:} Using the same density field creation procedure described in the power spectrum section, we calculate the matter bispectrum defined by, 
\begin{equation}
    \langle(\delta(\boldsymbol{k}_1)\delta(\boldsymbol{k}_2)\delta(\boldsymbol{k}_3)\rangle = (2\pi)^3 \delta^3_{\rm D}(\boldsymbol{k}_1+\boldsymbol{k}_2+\boldsymbol{k}_3)B(k_1,k_2,k_3)
\end{equation}
and reduced bispectrum,
\begin{equation}
   Q(k_1, k_2, k_3) \equiv\frac{B(k_1,k_2,k_3)}{P(k_1)P(k_2) + P(k_2)P(k_3) + P(k_3)P(k_1)}.
\end{equation}
We consider three different versions of the bispectrum; the equilateral configuration with $k_1=k_2=k_3$, the configuration obtained from varying the angle between two fixed-length wave vectors with $k_1 = 0.1$ Mpc$^{-1}h$ and $k_2 = 1.0$ Mpc$^{-1}h$, and the squeezed bispectrum with the parameter $S = 1- \frac{k_1}{k_2+k_3} = 1-\frac{k_1}{2k_2}$ where we have set $k_2 = k_3$.

In Fig. \ref{fig:Bk}, we calculate the bispectrum $B$ (top two rows) and the reduced bispectrum $Q$ (bottom two rows) in three different ways. In the left column, we show the equilateral bispectrum with $k$ ranging from $0.1$ up to $0.9$ Mpc$^{-1}h$ . At the largest scales investigated, with $k{\sim} 0.1$ Mpc$^{-1}h$, all cosmologies find agreement between the $N$-body and the emulator within $1\%$ for both the normal and reduced bispectrum, although as the scale decreases and $k \rightarrow 0.9$ Mpc$^{-1}h$, the emulator error tends to grow, up to ${\sim} 4\%$. For the reduced bispectrum, this error seems to grow monotonically with decreasing $\Omega_{\rm m}$, signalling that the emulator underestimates density at the scale of the haloes. The middle column again shows $B$ and $Q$, with $k_1 = 0.1$ Mpc$^{-1}h$ and $k_2 = 1.0$ Mpc$^{-1}h$ and $\theta \ \epsilon [0, \pi]$. The emulator's bispectrum is accurate within $3\%$ for all cosmologies across all values of $\theta$, and the reduced bispectrum shows higher accuracies within ${\sim} 1\%$. Finally, the last column shows the squeezed bispectrum. When configurations of this squeezed bispectrum, defined as $S = 1- \frac{k_1}{k_2+k_3}$, approach $1$ or $k_1 \ll k_2, k_3$, they can be considered responses of the local power spectrum to a long-wavelength density perturbation. Given the emulator's finite receptive field, and having only seen scales up to $1 \ h^{-1}$ Gpc during training, it is not guaranteed that the emulator will be accurate for scales beyond $1 \ h^{-1}$ Gpc. This squeezed configuration is a direct test of the emulator's performance outside of its field-of-view, where we set $k_2 = k_3 =0.1 \ h^{-1}$ Mpc and $k_1$ to range from $0.0024 \ h^{-1}$ Mpc (slightly larger than the fundamental wave number) up to $ 0.062  \ h^{-1}$ Mpc. The vertical dashed lines show the value of $S$ when $k_1$ is set to the field of view of the emulator, L$_{\rm FOV} \sim 200   \ h^{-1}$ Mpc, or $k_{\rm FOV} \sim 0.03  \ h^{-1}$ Mpc.
We find that the emulator performs extremely well for most squeezed configurations. Most notably, the emulator continues to achieve sub $2\%$ accuracies for most cosmologies on scales where $k_1$ is outside of the emulator's field-of-view, $k_1 < k_{\rm FOV}$ or $S>0.85$. This holds as $k_1$ approaches the largest scales ($k_1 {\sim} 0.006 \ h^{-1} $ Mpc) for most cosmologies, with the exception of high $\Omega_{\rm m}$ cosmologies where the error grows up to ${\sim} 5\%$.\\

\noindent \textbf{Wavelet Statistics:} The wavelet scattering transform has gained attention in cosmology as a tool to capture non-Gaussian information beyond traditional summary statistics \citep{Valogiannis_2022}. We use \texttt{Kymatio} to compute the wavelet scattering statistics \citep{kymatio}. To calculate the coefficients of this scattering transform, we first construct the density field of the particles following the prescription outlined in the power spectrum section. \texttt{Kymatio} then uses that density field to calculate the zeroth, first, and second order scattering coefficients by convolving the density field with wavelets and evaluating the moments. For more details on this calculation, see equations (1) and (2) of the \texttt{Kymatio} paper, \cite{kymatio}.

\begin{figure}
    \centering
    \includegraphics[width=\columnwidth]{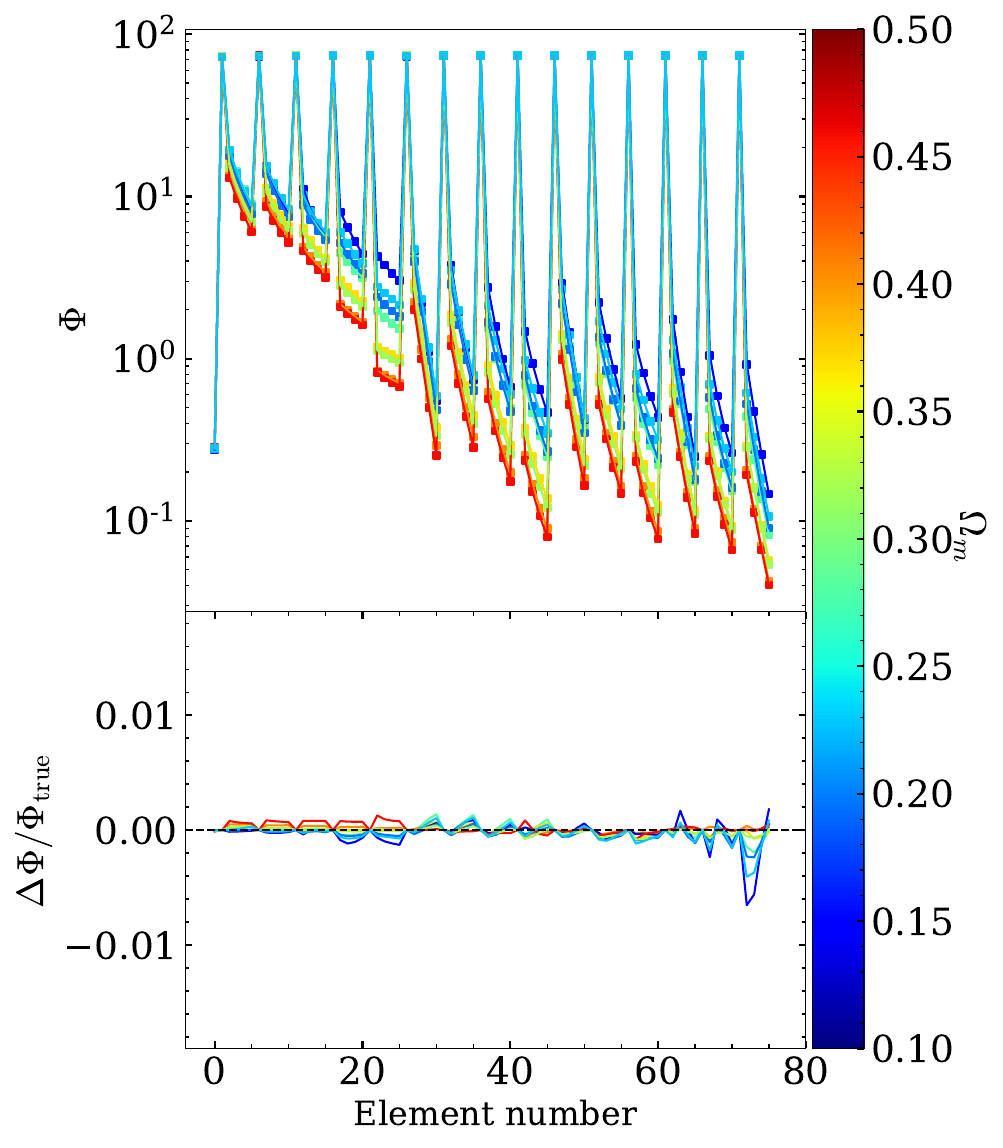}
    \caption{ We show the $N$-body (solid) and Emulator (square) outputs of the wavelet statistics applied to the particle fields.  Each colour corresponds to one value of $\Omega_{\rm m}$ in our eight cosmologies, with the value shown in the right colour bar. Each calculation was performed on a density field created with the CIC mass assignment scheme on a grid of $1536^3$ cells. }
    \label{fig:wavelet}
\end{figure}

We set the number of octaves to $J=4$ and the number of scales per octave to $Q=4$. We evaluate the wavelet coefficients for integral power $p=0.8$. The wavelet statistics are shown in Fig. \ref{fig:wavelet}. We find that the emulator and the $N$-body simulation agree within $0.5\%$ for every element number.

\subsection{Haloes}

\begin{figure}
    \centering
    \includegraphics[width=\columnwidth]{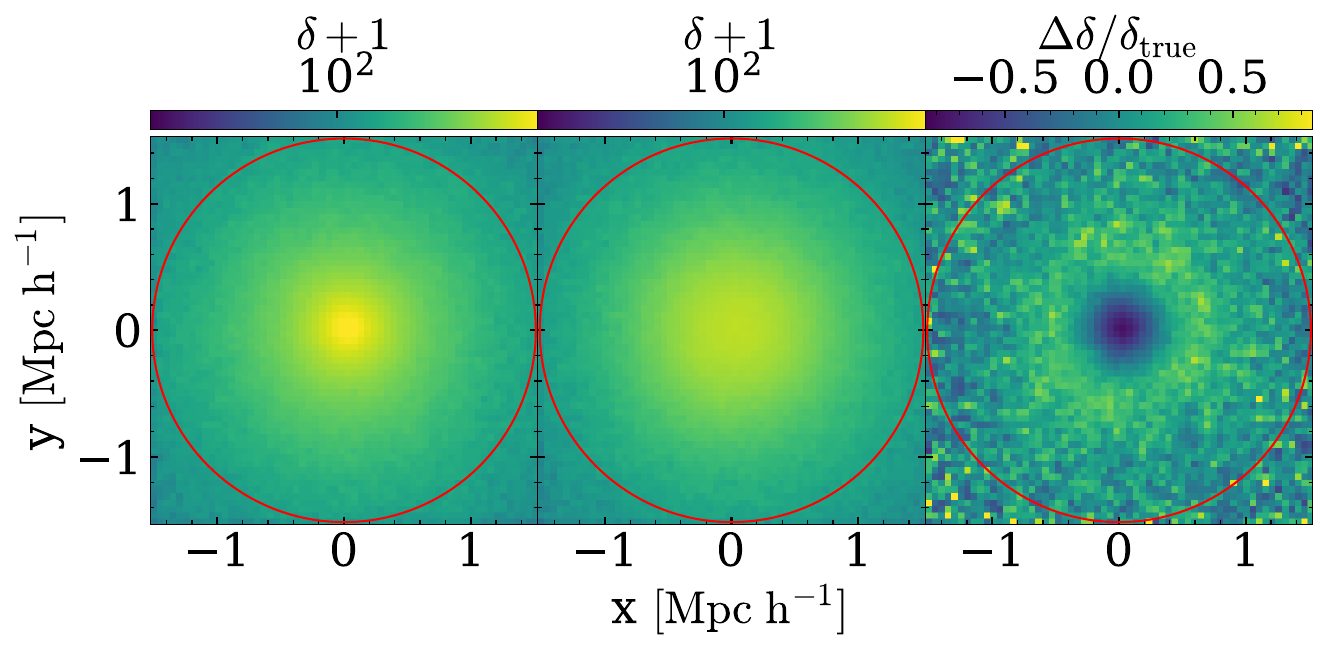}
    \caption{The stacked profiles for 500 haloes with mass closest to $\log_{10}(M_{\rm{target}}/$M$_\odot) = 14.5$ for the fourth cosmology in Table.~\ref{tab:table1}, setting $\Omega_{\rm m} = 0.1429$, $h=0.5049$, with a particle mass of $m_{\rm p}=2.95 {\times} 10^{11} \ h^{-1}{\rm M}_\odot$, meaning each halo has roughly 1,000 particles. The left panel shows the average density for 500 haloes in the $N$-body simulation, the middle shows the same for the emulator, and the right panel shows the fractional residuals (subtracting $N$-body from emulator). The width of panel is set to the virial diameter of the target halo, which is shown in red. We use a CIC method to count the average number of particles in each cell, using a cell resolution of $0.025$ $h^{-1}$Mpc, and averaging over the z-axis in a slice that is $30\%$ of the virial diameter.}
    \label{fig:stacked}
\end{figure}

\textbf{Halo Stacking:} We show an example of the stacked profiles for 500 haloes with a target mass of $\log_{10}(M_{\rm{target}}) = 14.5$ for a single cosmology in Fig. \ref{fig:stacked}. The cosmology sets $\Omega_{\rm m} = 0.1429$, $h=0.5049$ with a particle mass of $m_{\rm p}=2.95 \times 10^{11} \ h^{-1}$ M$_\odot$, meaning each halo has roughly 1,000 particles. The left panel shows the average CIC density for 500 haloes in the $N$-body, the middle panel shows the same results but for the emulator, and the right panel shows the fractional residuals (subtracting $N$-body from emulator and normalizing to the $N$-body). The width of the panel is set to the virial diameter of the halo with the given target mass for this cosmology, with the haloes virial radii shown in red. We use a CIC mass assignment scheme, with a resolution of ${\sim} 0.025$ $h^{-1}$Mpc per cell. We average over the z-axis, after selecting a slice that is $30\%$ the virial diameter and centred on the halo. Colour-bars indicate the over-density or the fractional residual.

The particles in the haloes of the $N$-body simulation (left) are much more concentrated near the centre of the halo. While the emulator can match halo masses within a few percent of the $N$-body simulation, it fails to resolve the highly non-linear density profile in the haloes' interior. Instead, it smooths out this density profile, distributing particles that should be near the centre closer to the outer edge of the halo. This is similar to what is found in other field level emulators (e.g. see Fig. C4 of \citet{Doeser2024}).

\begin{figure*}
    \centering
    \includegraphics[width=2\columnwidth]{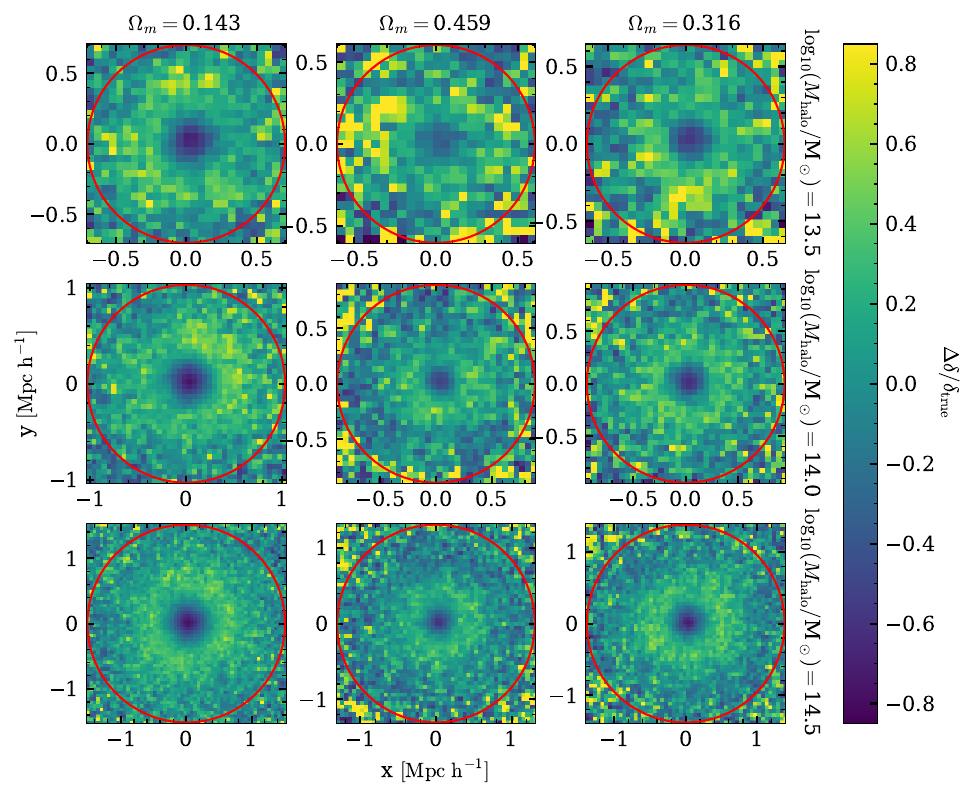}
    \caption{The fractional residuals of the stacked profiles for three target cosmologies, with the left, middle and right columns showing $\Omega_{\rm m} =0.143$, $0.315$, and $0.459$ respectively. Each row corresponds to a different target mass, with target mass shown on the right, where  $\log_{10}(M_{\rm{target}}/$M$_\odot) =13.5$, $14$ and $14.5$. The width of each panel is set to the virial diameter of the target halo, which is shown in red. We use a CIC method to count the average number of particles in each cell, using a cell resolution of $0.025$ $h^{-1}$Mpc, and averaging over the z-axis in a slice that is $30\%$ of the virial diameter.}
    \label{fig:stacked_residuals}
\end{figure*}

We show the fractional residual of 500 haloes in both the $N$-body and the emulator for a combination of 3 target masses and 3 different cosmologies in Fig. \ref{fig:stacked_residuals}. To calculate these fractional residuals, we follow the same procedure described previously.  Each column fixes a cosmology, with the value of $\Omega_{\rm m}$ shown above each column. Each row fixes a target mass, with $\log(M_{\rm halo}/{\rm M}_\odot)$ shown to the right of each row. With the goal of a sampling a wide range of $\Omega_{\rm m}$ values, we show cosmology 3, 5, and 6 from Table \ref{tab:table1}, which set $\Omega_{\rm m}$ to be $0.1429$, $0.4592$, and $0.3157$ respectively. Our target masses include $\log(M_{\rm halo}/{\rm M}_\odot) \ \epsilon \{13.5, 14.0, 14.5\}$. The virial radius for the given target mass and cosmology is shown in red.

For these three cosmologies and target masses, the emulator underestimates the concentration of particles near the centre of the halo by more than $50\%$, instead distributing these particles at roughly a half virial radius away from the centre. As $\Omega_{\rm m}$ decreases, the error tends to grow. This is expected, as a decrease in matter results in shallower potential wells, allowing particles to easily drift outward in these haloes. This means the initial displacement fields for the particles tend to be larger for lower $\Omega_{\rm m}$ cosmologies, or the PDF of initial displacements has a longer tail. As it is well known that neural networks underestimate extreme values due to their convolution operations being incapable of outputting large numbers (\citealt{Ting_2024}), it is reasonable that the stacked profiles for the haloes (and other summary statistics) are worse for lower $\Omega_{\rm m}$. We note that, for a small additional computational cost, there exists a method which could correct these errors. The COmoving Computer Acceleration (COCA) framework may be able to correct these errors \citep{Bartlett_2024}. By solving the correct physical equations in an emulated frame of reference, they find that COCA reduces emulation errors in particle trajectories and still reduces the computational budget compared to traditional $N$-body simulations alone.\\

\begin{figure}
    \centering
    \includegraphics[width=\columnwidth]{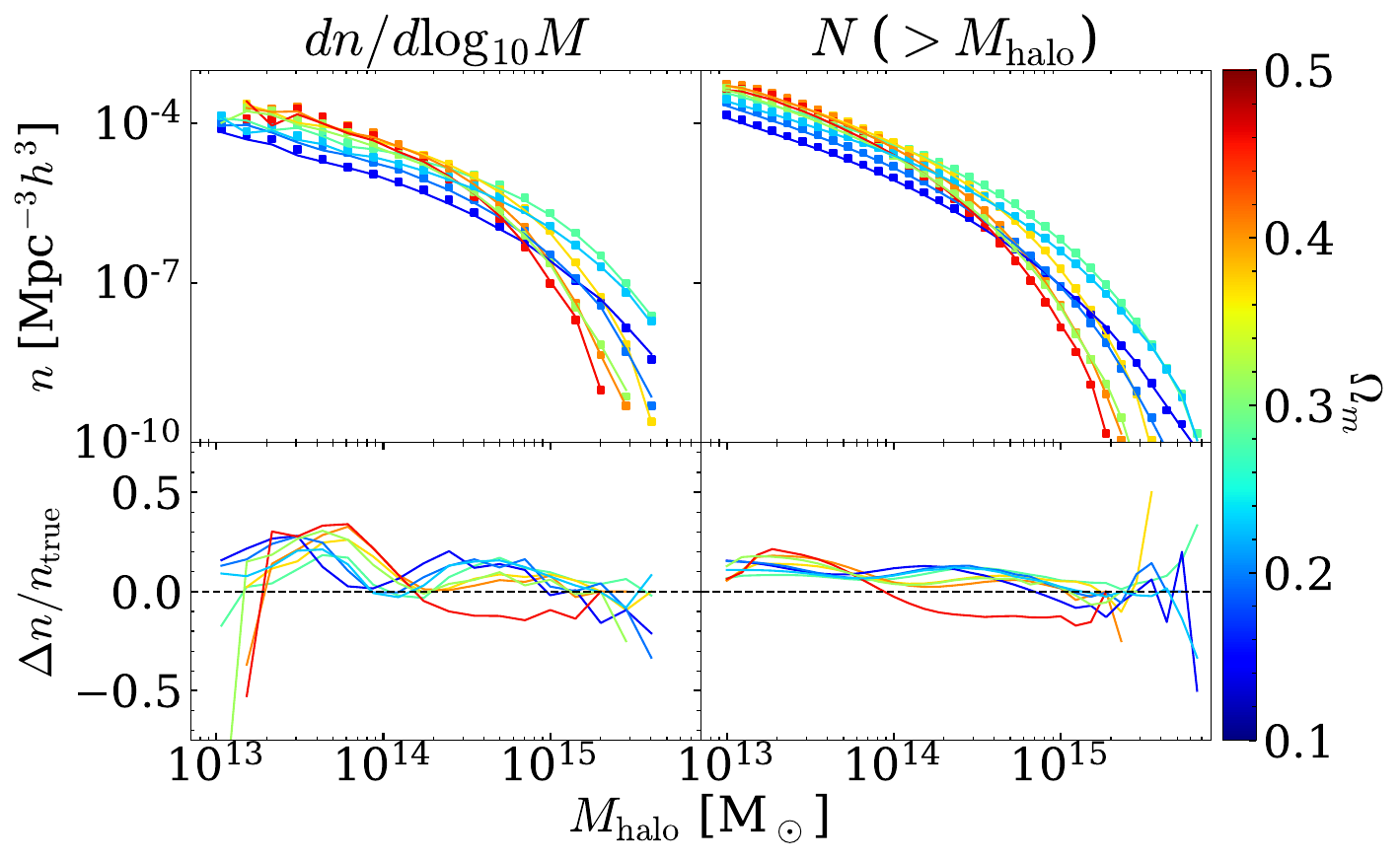}
    \caption{We show the halo mass function (left) and the cumulative mass function (right). $N$-body simulations are shown in solid and Emulator outputs are marked by squares. The bottom panels show the fractional residuals, subtracting the $N$-body results from the emulator. Each colour corresponds to one value of $\Omega_{\rm m}$ in our eight cosmologies, with the value shown in the right colour bar. }
    \label{fig:hmf}
\end{figure}

\noindent \textbf{HMF:} The halo mass function and cumulative mass function are shown in Fig. \ref{fig:hmf}. We show the absolute values in the top row and their fractional error in the bottom row. The emulator distributes dark matter in a way that tends to slightly overestimate halo mass, leading to surplus of haloes above our cut of $10^{13} h^{-1}{\rm M}_\odot$ in every cosmology. The emulator has roughly $10\%$ to $20\%$ more haloes above that cut (shown in the bottom right panel), though this error generally decreases with increasing halo mass, with the exception of the high $\Omega_{\rm m}$ cosmology, where the emulator underestimates the abundance of haloes with mass $M_{\rm halo} > 10^{14} {\rm M}_\odot$. We note that there exists several techniques which correct the errors in halo statistics that could improve these results in post-processing for a small additional computational expense \citep{Izard_2015, Dai_2019, Wu_2024}.\\

\begin{figure*}
    \centering
    \includegraphics[width=\columnwidth]{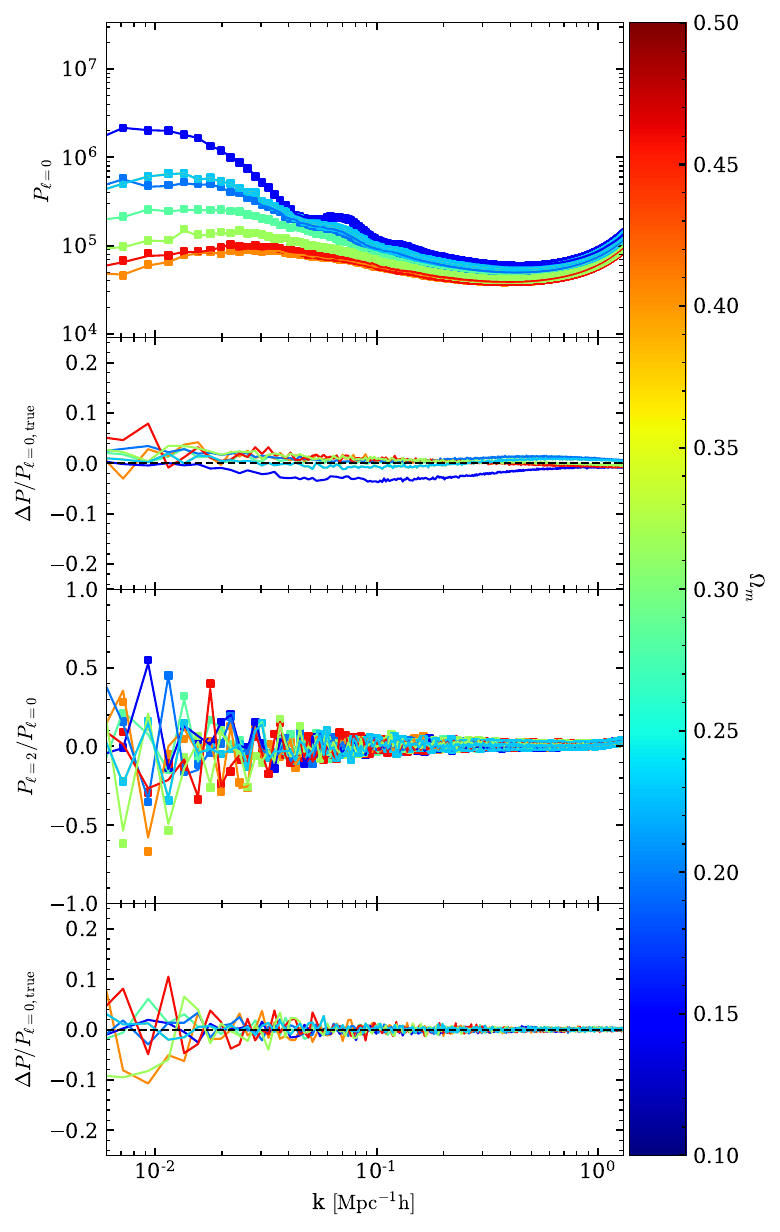}
    \includegraphics[width=\columnwidth]{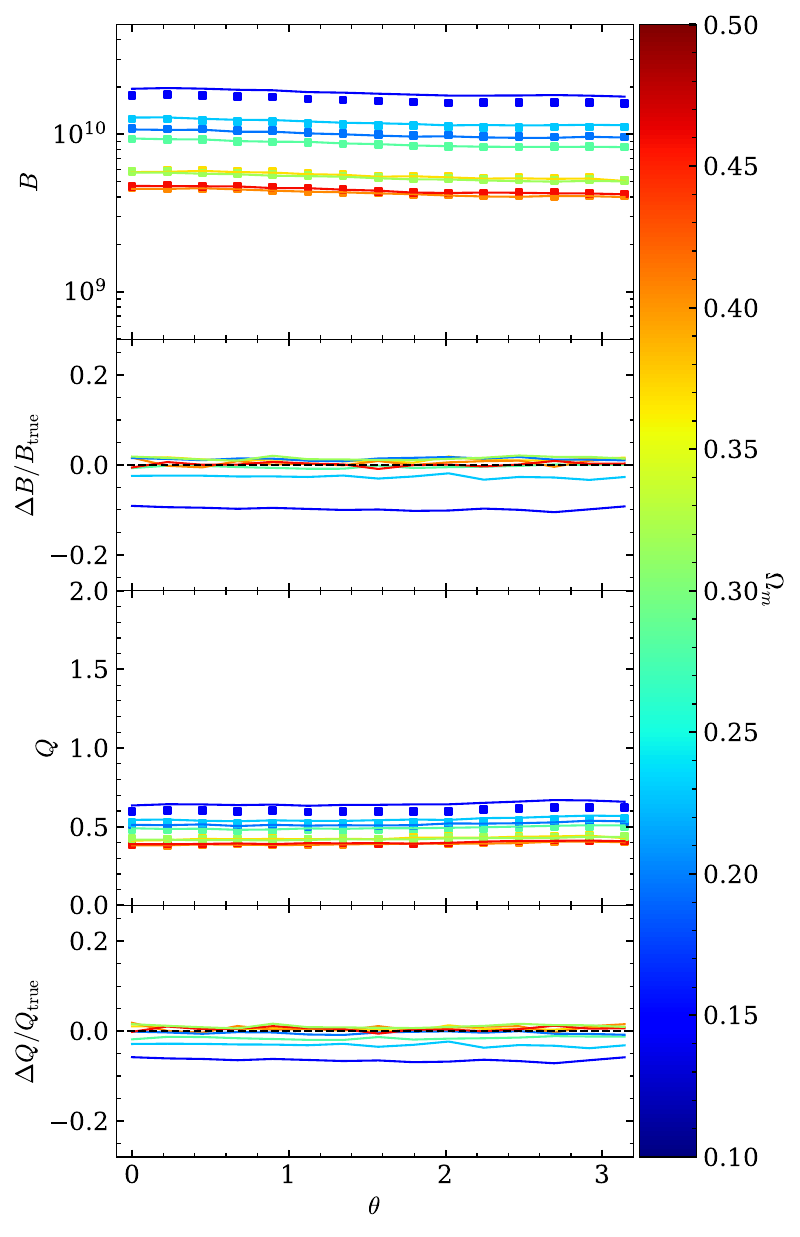}

    \caption{Left: A comparison of the monopole and quadrupole moments of the mass weighted power spectrum for the \texttt{Rockstar} haloes in real space. Right: The bispectrum and reduced bispectrum for the haloes in real space. We construct a density field using the most massive $10^6$ haloes with a mass weight of $w = (M_{\rm halo}/(10^{14} {\rm M}_\odot))^{\alpha}$, setting $\alpha =0.7$, on a grid of $1536^3$ cells. Each colour corresponds to one value of $\Omega_{\rm m}$ in our eight cosmologies, with the value shown in the right colour bar.}
    \label{fig:pk_bk_haloes}
\end{figure*}

\noindent \textbf{Power Spectrum:} We compute the power spectrum of the haloes in the same way that we compute the power spectrum of the particles, with the addition of weighing each halo following \citet{Pandey_2024}. We select the most massive million haloes, and weight them according to $w = (M_{\rm halo}/(10^{14} {\rm M}_\odot)^{\alpha}$ with $\alpha =0.7$, which mimics the SDSS-CMASS halo occupation distribution weighting \citep{Reid_2014, Hahn_2023}. We then deposit these weighted haloes onto a $1536^3$ over-density field and use \texttt{Pylians} to compute the power spectrum of the haloes in real space. The results for the monopole, quadrupole, and their fractional residuals are shown in the left panel of Fig. \ref{fig:pk_bk_haloes}.

For the power spectrum monopole, we find excellent agreement for most cosmologies, typically within $<3\%$ across most values of $k$. Though, for the cosmology with the smallest value of $\Omega_{\rm m} =0.1429 $, the power spectrum monopole is underestimated in the emulator by nearly $6\%$ for $k {\sim} 0.1 $ Mpc$^{-1}h$. It is reasonable that the error would increase with decreasing $\Omega_{\rm m}$, as the smallest halo of our selection from the million most massive would be much less massive than the same halo for a cosmology with a larger $\Omega_{\rm m}$, and we have demonstrated that the error in the haloes tends to increase with decreasing mass and the fewer number of particles they contain. For the cosmology with the smallest value of $\Omega_{\rm m}$, $\Omega_{\rm m}=0.1429$, the millionth most massive halo has a mass of $3.132 \times 10^{13} \ h^{-1}{\rm M}_\odot$, or ${\sim} 100$ particles, which is approaching the minimum number of particles required to reliably calculate halo statistics \citep{Mansfield_2021}, coupled with the effect regarding the emulator performing worse for lower values of $\Omega_{\rm m}$. For the quadrupole, errors are the largest at the larger scales ($k {\sim} 0.01$ Mpc$^{-1}h$) with an error of up to $10\%$, though we achieve agreement within ${\sim} 1\%$ across all cosmologies for smaller scales.\\

\noindent \textbf{Bispectrum:} Using the same mesh size and weights, we compute the bispectrum of the haloes, shown in the right panel of Fig. \ref{fig:pk_bk_haloes}. We calculate the bispectrum (and reduced bispectrum) with $k_1 = 0.1$ and $k_2 = 1.0$ Mpc$^{-1}h$. Most cosmologies see excellent agreement between the emulator and the $N$-body (within $3\%$), with the exception of the cosmology with the smallest $\Omega_{\rm m}$, which has a consistent ${\sim} 10\%$ error across all $\theta$. The error is roughly the same for the reduced bispectrum, though this error is dampened. As noted in the HMF section, the errors found in the statistics of the haloes could be improved with halo mass corrections, which would improve the weights applied to the power spectrum and bispectrum.

\section{Conclusion}
\label{sec:conclusion}
In this work, we carry out the first comprehensive test of an accelerated forward model at volumes matching next-generation surveys. We have tested a field level $N$-body emulator on eight $(3\ h^{-1}{\rm Gpc})^3$ simulations with cosmological parameters sampled from a Sobol sequence outlined in \S~\ref{sec:methods}. Our results at these larger scales are similar to the results of \citet{Jamieson_2023}, where we achieve high accuracy (within ${\sim}$1\% level accuracy) in the power spectrum, bispectrum, and WST of the particle fields at scales ranging from as low as $k{\sim} 0.0024$ $h^{-1}$Mpc down to the small non-linear scale of $k{\sim} 1.0$ $h^{-1}$Mpc. For the power spectrum, in real space, the emulator is within $1\%$ of the $N$-body at all scales explored. For the RSD power spectrum, the emulator achieves a similar agreement for $k \leq 0.2$ $h^{-1}$Mpc, though results diverge at smaller scales. For all cosmologies explored, the equilateral, $\theta$-dependent, and squeezed bispectrum (and their derived reduced bispectrum) all achieve sub $3\%$ accuracy. Notably, the squeezed bispectrum shows that the emulator performs extremely well for the most squeezed configurations when $k_1 << k_2 {\sim} k_3$, signalling that it recovers the large scale modes, even those well outside the emulator's field-of-view.

We have also measured the fractional residuals of stacked profiles for haloes generated through \texttt{Rockstar}, as well as the HMF, the power spectrum and the bispectrum for these haloes. The fractional residuals of the stacked profiles indicate that the emulator is able to approximately mimic the haloes in an $N$-body simulation, though it has difficulty resolving their non-linear centres. We have highlighted that this error grows with decreasing $\Omega_{\rm m}$, which can be attributed to the neural networks inability to estimate the extreme displacements of particles in smaller $\Omega_{\rm m}$ cosmologies. We note that this error is correctable, using the method outlined in \citet{Bartlett_2024}. The halo mass function suggests that the emulator systematically overestimates the mass of haloes $\gtrsim 10\times 10^{13} h^{-1} {\rm M}_\odot$, resulting in up to $20\%$ more haloes above this threshold for the emulator. However, there exists techniques to correct these halo mass errors \citep{Izard_2015, Dai_2019, Wu_2024}.

The mass weighted power spectrum and bispectrum results show a similar trend. The emulator's power spectrum monopole and quadrupole is in excellent agreement with the $N$-body for most $k$ values (typically $< 2 \%$), as well as the bispectrum and reduced bispectrum agreeing within ${\sim} 5\%$ for most cosmologies. The largest errors come from the cosmologies with the smallest $\Omega_{\rm m}$ values. Using the newer version of this emulator that was released while writing this manuscript \citep{Jamieson2024}, which encodes a redshift dependence and improves small-scale velocity estimates, would yield equivalent or slightly better results than we've found here. Additionally, all of the tests carried out here are at redshift $z=0$, though at $z=0.5$, which is the interest of modern LSS surveys, the new redshift-dependent model is much more accurate and errors at $k{\sim} 1.0$ $h^{-1}$Mpc scales would be $<1\%$. However, we have achieved our primary goal of extending an emulator to volumes that are larger than it was originally trained on, and have found that this extension does not create significant errors. We have also analysed the haloes recovered from this large scale application, finding the power spectrum and bispectrum of the haloes agree with their $N$-body counterparts within percent-level accuracies across most scales, comparable to mock halo catalogue generators \citep{Pandey_2024}. Overall, the emulator has proved to be a promising, powerful tool for expediting non-linear simulations at large $(3\ h^{-1}{\rm Gpc})^3$ volumes matching the observable volumes of next-generation surveys.

\section*{Acknowledgements}
We thank the CCA at the Flatiron Institute where some of this research was carried out. This work is supported by the Simons Collaboration on "Learning the Universe".  Simulations and data analysis were performed with NSF’s ACCESS allocation AST-140041 on the Stampede3 resource. GLB acknowledges support from the NSF (AST-2108470 and AST-2307419, ACCESS), a NASA TCAN award, and the Simons Foundation. The freely available plotting library matplotlib (\citealt{Hunter_2007}) was used to construct the plots in this paper. 

\section*{Data Availability}
The trained model parameters are available at \href{https://github.com/dsjamieson/map2map_emu}{map2map\_emu}. This model uses \href{https://github.com/eelregit/map2map}{map2map}, which is a neural network framework based on \texttt{PyTorch} \citep{pytorch}. The code used to run the emulator and analyse the data is available at this \href{https://github.com/mscoggs/ltu_emulator_halos}{github repository}. All other data will be shared on reasonable request to the corresponding author.

\bibliographystyle{mnras}
\bibliography{references} 
\bsp
\label{lastpage}
\end{document}